\documentclass[preprint2,longnamesfirst]{aastex}

\newcommand{\Ma}{\mathop{\rm Myr\,}\nolimits}
\newcommand{\Ga}{\mathop{\rm Gyr\,}\nolimits}

\newcommand{\kpc}{\mathop{\rm kpc\,}\nolimits}
\newcommand{\Mpc}{\mathop{\rm Mpc\,}\nolimits}
\newcommand{\kps}{\mathop{\rm km \ s^{-1}\,}\nolimits}
\newcommand{\Mach}{\mathop{\mathcal{M\,}\,}\nolimits}

\newcommand{\erg}{\mathop{\rm erg\,}\nolimits}
\newcommand{\keV}{\mathop{\rm keV\,}\nolimits}
\newcommand{\K}{\mathop{\rm K\,}\nolimits}

\newcommand{\Msun}{\mathop{\rm M_{\odot}\,}\nolimits}

\newcommand{\fig}{Fig.~\ref}

\newcommand{\Fig}{Figure~\ref}

\newcommand{\sect}{Sec.~\ref}

\newcommand{\rosat}{{\it ROSAT}}

\newcommand{\chandra}{{\it Chandra}}

\newcommand{\expd}[1]{\times 10^{#1}}

\newcommand{\fraction}[2]{\mbox{\scriptsize$^{{#1}\!}/_{\!{#2}}$}}

\ifx\undefined\onequarter
 \newcommand{\onequarter}{\fraction{1}{4}}
\else
 \renewcommand{\onequarter}{\fraction{1}{4}}
\fi
\ifx\undefined\onefourth
 \newcommand{\onefourth}{\onequarter}
\else
 \renewcommand{\onefourth}{\onequarter}
\fi
\ifx\undefined\onethird
 \newcommand{\onethird}{\fraction{1}{3}}
\else
 \renewcommand{\onethird}{\fraction{1}{3}}
\fi
\ifx\undefined\onehalf
 \newcommand{\onehalf}{\fraction{1}{2}}
\else
 \renewcommand{\onehalf}{\fraction{1}{2}}
\fi
\ifx\undefined\twothirds
 \newcommand{\twothirds}{\fraction{2}{3}}
\else
 \renewcommand{\twothirds}{\fraction{2}{3}}
\fi
\ifx\undefined\threequarters
 \newcommand{\threequarters}{\fraction{3}{4}}
\else
 \renewcommand{\threequarters}{\fraction{3}{4}}
\fi

\ifx\undefined\onetenth
 \newcommand{\onetenth}{\fraction{1}{10}}
\else
 \renewcommand{\onetenth}{\fraction{1}{10}}
\fi

\newcommand{\figscaleone}{1}

\begin{document}

\title{Cluster mergers, core oscillations, and cold fronts}
\shorttitle{Cluster mergers oscillations and cold fronts} 
\author{Eric R. Tittley\altaffilmark{1} and Mark Henriksen}
\affil{Joint Center for Astrophysics, University of Maryland, Baltimore County}
\affil{Baltimore, MD 21250, USA}
\altaffiltext{1}{present address:
                 Institute for Astrophysics, University of Edinburgh,
                 Blackford Hill, Edinburgh EH9 3HJ, UK}
\email{ert@roe.ac.uk}
\email{mark@jca.umbc.edu}
\shortauthors{TITTLEY \& HENRIKSEN}

\shortcites{Markevitch00,Mazzotta01}

\begin{abstract}
We use numerical simulations with hydrodynamics to demonstrate that a class of cold fronts in galaxy clusters can be attributed to oscillations of the dark matter distribution.  The oscillations are initiated by the off-axis passage of a low-mass substructure.  From the simulations, we derive three observable morphological features indicative of oscillations: 1) The existence of compressed isophotes; 2) The regions of compression must be alternate (opposite and staggered) and lie on an axis passing through the center of the cluster; 3) The gradient of each compression region must pass through the center of the cluster.  Four of six clusters reported in the literature to have cold fronts have morphologies consistent with the presence of oscillations.  The clusters with oscillations are A496, A1795, A2142, and RX J1720.1+2638.  Galaxy clusters A2256 and A3667 are not consistent so the cold fronts are interpreted as group remnants.  The oscillations may be able to provide sufficient energy to solve the cooling-flow problem and, importantly,  provide it over an extended duration.
\end{abstract}

\section{Introduction}
\label{sec.Intro}

The high spatial resolution of the \chandra\ x-ray telescope has led to the discovery of galaxy cluster phenomena neither previously seen
by x-ray telescopes nor previously predicted by numerical simulations.
Among these are cold fronts, a term coined in \citet{VMM01b} and observed in many clusters including the prototypes Abell 2142 \citep{Markevitch00}, Abell 3667 \citep{VMM01b}, and RX J1720.1+2638 \citep{Mazzotta01} as well as Abell 1795 \citep{MVM01}, Abell 2256 \citep{SMMV02}, and Abell 496 \citep{DW03}.

Cold fronts share the following two characteristics: 1) a steep, non-axisymmetric x-ray flux gradient and 2) a steep temperature gradient with the cooler gas on the denser side in approximate pressure equilibrium.  The density and temperature gradients at the front are sufficiently steep to allow them to be modeled as discontinuities, within the resolution limits of the detector.

Cold fronts are not shock fronts.  \citet{Markevitch00} estimates $\Mach < \onehalf$ for the front in A2142.  For A3667, \citet{VMM01b} estimate $\Mach = 1 \pm 0.2$ from the ratio of the pressure on either side of the front.  The same method applied to RX J1720.1+2638 gives $\Mach = 0.4^{+0.7}_{-0.4} $\citep{Mazzotta01}. 

Steep, non-axisymmetric luminosity gradients were observed by the \rosat\ x-ray telescope but misinterpreted as shock fronts \citep{MSV99}.  Indeed, non-axisymmetric luminosity gradients had been noted in many clusters prior to
\chandra\ \citep{MFG93, MEFG95, BT96}.

The presently postulated explanation for cold fronts is that they are the low-entropy remnant of a recently merged substructure \citep{Markevitch00}.   The low-entropy gas is in the process of, or has been stripped via ram pressure from the merging substructure.  The temperature discontinuity can arise from either the parent halo suffering shock heating, or the low-entropy core cooling adiabatically as it expands in response to the loss of its potential well.  A combination of factors is possible.

The subcluster-remnant scenario begs the question: where is the parent cluster's low-entropy core? The cool dense gas of the cold front is the remnant of the group.  Radiative cooling is more efficient in the dense centers of massive clusters.  This should lead to a large reservoir of low-entropy gas.  This cool gas is observed in large clusters, though not in the abundance anticipated by the heuristic argument given which is the classic ``cooling flow'' problem.

This paper proposes cold fronts are a symptom of more than one phenomenon.
The previously postulated explanation that the front is due to gas from merging object is valid only for one class of fronts which we call here the ``remnant class''.
We suggest another class of fronts in which the front is due to oscillations of the core after the merger.  The significant difference is the cool gas is original, central gas from the parent halo.  This class is called the ``oscillation class'' in the rest of this paper.  This model shares many traits with the model proposed in \citet{MVM01}. In their model, the cool gas is sloshing in the gravitational potential of the cluster.  The difference is that in the model presented here, it is the cluster gravitational potential that is oscillating, not the gas within the potential.  The dark matter is moving {\it en masse} in an oscillatory fashion, carrying the bulk of the gravitational potential with it.  The displacement of the gravitational potential induces motion in the gas as well.  The model of \citet{MVM01} requires a non-gravitational perturbation to the gas in the core.  Such a perturbation can be provided by an AGN jet or large-scale turbulent motion induced by matter in falling along a filament \citep{NK03}.  Hence, their model potentially represents another class of cold fronts.

\citet{Mazzotta01} suggests a model which is a hybrid of the remnant and oscillation classes: a subcluster forms co-spatially with a larger halo but offset sufficiently such that the subcluster oscillates within the center of a larger halo.  This model differs from the remnant class in that the cool gas is never stripped from its dark matter halo and differs from the oscillation class in that it is the subcluster that is oscillating, not the primary.

There are two advantages to the oscillation model presented here.
\begin{enumerate}
\item The oscillations are able to persist long after the passage of the
	merging structure.
\item Major mergers are not required to instigate oscillations.  More
	common high-mass-ratio interactions are sufficient.
\end{enumerate}
Together, these two advantages predict that cold fronts should be common, and not unique to clusters exhibiting recent major merger activity. Evidence supporting this prediction has been reported.  \citet{MVF02} find 25 of 37 clusters selected from \rosat\ images as relaxed (at least in morphology) have brightness edges similar to those in typical cold front clusters.  Such a large fraction of clusters with cold fronts is in conflict with expectations.
\citet{FSNY02}, using an analytical model of the mass distribution function, find that $\onethird$ of clusters should have large subhaloes.  Not all subhaloes should exhibit cold fronts, implying that less than $\onethird$ of clusters should have cold fronts.

In a set of simulations of preheated galaxy clusters, \citet{BEM02} found one cluster with a structure reminiscent of cold fronts.  The cluster was undergoing a recent merger of comparably-massed objects, hence belongs to the remnant class. They simulated 68 clusters and found only one instance of a cold front,
suggesting that remnant-class cold fronts are rare.

\citet{NK03} performed simulations of cold dark matter cluster mergers specifically designed to study cold fronts of the remnant class.  They verify the transient nature of cold fronts around the substructure remnant.

Chemical gradients provide a further constraint to test the model: in particular, the absence of a gradient.  The absence of a gradient is difficult to explain if the cool, dense gas of the cold front is a remnant of an alien structure.  However, the oscillation model predicts no chemical gradient across the cold front unless a chemical gradient existed prior to and independently of the perturbing interaction.  The model predicts density gradients in regions in which the local gas is compressed.  The compression increases the gradient of any chemical variation that already exists, but does not in itself inject gas with a dissimilar chemical history.
The absence of a gradient in the A496 cold front is reported by \citet{DW03}. Mergers have been postulated to destroy chemical gradients \citep{DM01}; turbulence generated by the merger efficiently mixes the gas.  However, the remnant would necessarily be destroyed by the mixing.

Of the well-studied cold front clusters, we argue below that the fronts in A2256 and A3667 are members of the remnant class, while RX J1720.1+2638, A496, A1795, and A2142 are members of the oscillation class.

The paper is laid out as follows.  In \sect{sec.Simulations}, the methodology and initial conditions of the simulations are described.  The results of the simulations are described in \sect{sec.SimulationResults} and observational consequences are derived.  With the observational criteria, in \sect{sec.Comparison} six cold front clusters from the literature are examined in detail and classified as either remnant or oscillation class.  The possibility that the energy is sufficient to explain the cooling flow problem is explored in \sect{sec.Energy}. We summarize the results in \sect{sec.Summary}.

\section{Simulations}
\label{sec.Simulations}

We used a series of numerical simulations to explore the effects of a merger on the core of a cluster.  Both the collisionless dark matter and the gas were evolved using the hydro/N-body code, Hydra \citep{CTP95}, which models the
hydrodynamics using smoothed particle hydrodynamics (SPH).  The initial clusters and groups were extracted from a cosmological simulation of a $(40 \Mpc)^3$ volume.  Substructure in the haloes was maintained.  This provides for a more realistic state for the cluster and group haloes.
The cluster and group were evolved in isolated volumes for a few crossing times before combining.  The short evolution allows artifacts due to the abrupt change in the boundary conditions and local tidal field to dissipate.

{\bf Simulation A} models a high mass-ratio interaction.  A group of mass $4.7\expd{13} \Msun$ was collided with a cluster of mass $1.4\expd{15} \Msun$ for a mass ratio of 30:1.  The cluster and group were extracted from a CDM simulation.  The mass resolution of the cluster and group was boosted by a factor of two.  The cluster was modeled by 371 678 particles divided 57:43 dark matter to gas, a ratio set by evolution within the parent CDM simulation.  The group was modeled by 7940 particles in a dark matter to gas ratio of 62:38.  The dark matter particles had masses 13 times that of the gas particles.  The group was placed 4.6 Mpc from the cluster and given an initial velocity of $500 \kps$. This is a bound state; the bound/unbound transition is $1600 \kps$.  The initial impact parameter (defined as the separation vector of closest approach in the absence of gravitational acceleration) was 1 Mpc, leading to a closest approach of 110 kpc.  There was no radiative cooling in Simulation A.
The cluster-group system was evolved for an elapsed simulation time of $4 \Ga$.

To boost the resolution of Simulation A, each particle was split into two and distributed randomly within a local radius in which there were originally 32 particles.  For dark matter particles, the particles were given random velocities such that the total momentum, kinetic energy, and local velocity dispersion (determined from the 32 closest particles) of the original particle were conserved.

{\bf Simulation B} models an intermediate mass-ratio interaction.
A group of mass $4.3\expd{13} \Msun$ was collided with a cluster of mass $4.2\expd{14}\Msun$ for a mass ration of 10:1.
The cluster and group were extracted from a $\Lambda$-CDM simulation ($\Omega_m=0.3$, $\Lambda=0.7$). The simulation used 233 461 particles split almost 50/50 into dark matter and gas.  The ratio of the masses of the dark matter to gas was 5.7:1. The group was placed 6.2 Mpc from the cluster and given an initial velocity of $500 \kps$.  This is a bound state.  The bound/unbound transition is $763 \kps$.  The initial impact parameter was 1.5 Mpc, leading to a closest approach of 356 kpc. In contrast to Simulation A, radiative cooling was included.
The cluster-group system was evolved for an elapsed simulation time of $20 \Ga$.

\section{Simulation results}
\label{sec.SimulationResults}

\begin{figure}
\epsscale{\figscaleone}
\plotone{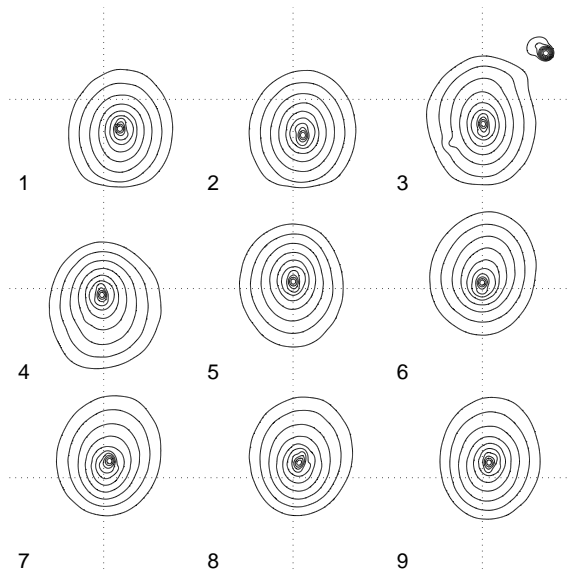}
\caption{
Oscillations induced by the passage of a merging sub-cluster in Simulation A. The contour lines
map projected gas mass. Contour lines are spaced by a factor of $\sqrt 2$. The
dotted lines give a constant frame of reference. Frames are $50 \Ma$ apart.
The scale of each box is 200 kpc.
}
\label{fig.Oscillations}
\end{figure}

Oscillations in the core are induced by the passage of the group. The core
moves with respect to the outer halo which, in turn, begins to move with
respect to the core.  The details are illustrated in \fig{fig.Oscillations}
which plots contours of equal projected mass at time intervals separated by $50
\Ma$ from Simulation A.  The group approaches from the left in frames 1 and 2 (group not seen), passes over the cluster and is deflected to the lower right where it continues out of the frames. Closest approach is 110 kpc in frame 3.  Compression of the contours are seen in frames 4, 6, and 7. In frame 4, the
compression is seen on the upper face, 16 kpc in the direction of the impact
radius and is due to the movement of the cluster center toward the initial
impact point. In frame 6, the cluster center is now stationary, but the halo
has moved upward, overshooting the core's displacement and causing compression
on the lower side 16 kpc away. The overshoot re-accelerates the core upward,
causing the core to compress material above itself, 8 kpc away, while the halo
is still moving upward, continuing the compression in the lower half at the
larger radius of 32 kpc. Compare frame 7 with Fig. 7 where there are not two
but three regions of compression laid out in a similar fashion. By frame 8, 300
Myr after the initial impact in frame 2, the core and halo have come to rest
centered on a point displaced 50 kpc above their initial position. Note that the
core oscillates perpendicularly to the initial velocity vector of the impactor.

At this point it is important to qualify what we are calling ``oscillation''.  No single isodensity contour oscillates in a periodic fashion (like a pendulum) with respect to the fixed reference frame.  Instead, motions are periodic only with respect to the cluster center, which is not a fixed reference frame.

From \fig{fig.Oscillations}, we estimate the duration of the oscillations is 250 Myr for Simulation A.  For the smaller mass ratio in Simulation B, the oscillations persist for at least 600 Myr but have dissipated by 1.2 Gyr.

Crucial to the model presented here are the following three observational consequences:
\begin{enumerate}
 \item There can be one, two, or three regions of compressed isodensity contours.
 \item The regions of compression must be alternate (opposite and staggered) of each other and lie on an axis which passes through the center of the cluster.
 \item The gradients of compressed isodensity must point through the center of the cluster.
\end{enumerate}
The remnant model cannot explain more than one region of compressed isodensity contours.  Coincidental mergers or separate substructures can produce two cold fronts. However, they need not be aligned along any preferred axis or be alternate.  The model presented by \citet{MVM01}, in which the gas is oscillating in a fixed potential set by the dark matter, predicts the third consequence, but does not explain the ringing which leads to the first two consequences.  Indeed, it is not obvious that a model in which the dark matter at the center is oscillating within a global potential, dragging the gas with it, will reproduce the ringing.  The alternate compression regions are due to the delayed response of the outer halo to the perturbation of the halo by the passing substructure. We will return to these three consequences in \sect{sec.Comparison}.

There exists the opportunity for confusion if the cluster-substructure interaction induces oscillation of the core and the substructure remains in, or subsequently falls back on the core to produce both a sloshing and merger-remnant cold front.  Such a cluster would have an merger-remnant cold front while the core of the primary cluster would be undergoing oscillations.  But these two cold-front phenomena are still distinct, even if they can exist in the same cluster at the same time.  An oscillation in a cluster with the merger remnant in the immediate vicinity has not been explored by these simulations.  Given that the presence and strength of the oscillation-type cold fronts is a function of time, it is likely that the dominant cold front class in a mixed scenario would be very sensitive to the stage of the merger.

There are other observational consequences which are not discussed here: isophotal twisting and multiple oscillation modes.
The possibility of isophotal twisting occurring exists, but is not explored in this paper since the effect is not as obvious.  However, we suspect that it would be a further probe of the merger geometry.  Multiple sets of oscillations would be initiated by multiple merger events (either separate objects or a single object on separate passes). If multiple oscillation modes are concurrent, the resultant morphology will be distorted making analysis problematic.

\begin{figure}
\epsscale{\figscaleone}
\plotone{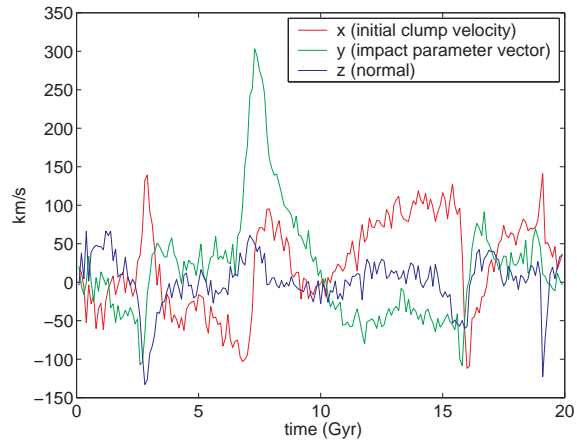}
\caption{
Core velocities in the x, y, and z directions for a 9:1 mass ratio, 300 kpc closest approach in Simulation B.
For reference, $c_s = 340 (T/\keV)^{\onehalf} \kps$.
}
\label{fig.CoreVelocities}
\end{figure}

The core velocity relative to the simulation volume for Simulation B is illustrated in \fig{fig.CoreVelocities}.  The strongest perturbation to the core of the primary halo is in the direction of the impact parameter, perpendicular to the initial group velocity vector.  The maximum velocity is $300 \kps$ which is less than the speed of sound of the cluster medium ($c_s \simeq 340 \kps$).

\begin{figure}
\epsscale{\figscaleone}
\plotone{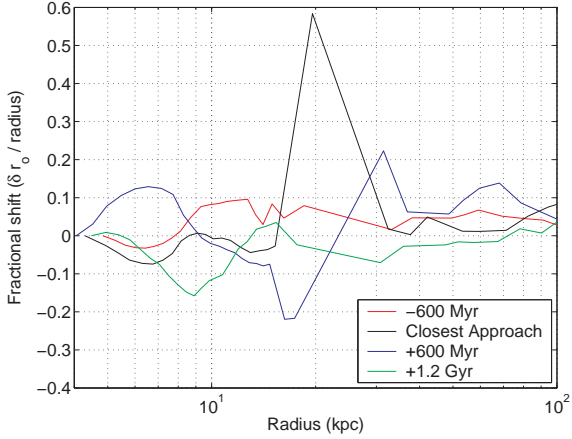}
\caption{
Fractional shifts in positions of isodensity spheres of varying radii at 4 times (600 Myr before closest approach, at closest approach, 600 Myr after closest approach, and 1.2 Gyr after closest approach).  Only the shift in the direction of the impact parameter is measured.  For Simulation B.
}
\label{fig.OscillationCurves}
\end{figure}

\fig{fig.OscillationCurves} illustrates the displacement of isodensity contours in response to the perturbation, with respect to the center of the primary cluster halo for Simulation B.  Plotted are the fractional shifts ($\delta r / r$) in the centers (along the impact radius vector) of isodensity surfaces of varying radii at different times. Closest approach of the perturbing group is 300 kpc.  Non-axisymmetric density gradients are created where the slopes of the curves are steepest.  Negative slopes indicate isodensity compression, positive slopes indicate isodensity stretching.  Before impact (600 Myr prior to closest approach), the maximum fractional variation is $< 10\%
$.  At closest approach, there is one large displacement of $\sim 60\%
$ for the isodensity contours at a radius of 20 kpc.  At 600 Myr after closest approach, there are three regions of isodensity contour compression: at 8 kpc, 35 kpc, and at $> 80$ kpc. By 1.2 Gyr, there is only a compressed gradient left in the center (6 to 8 kpc) and at large radii ($>200$ kpc).

\begin{figure}
\epsscale{\figscaleone}
\plotone{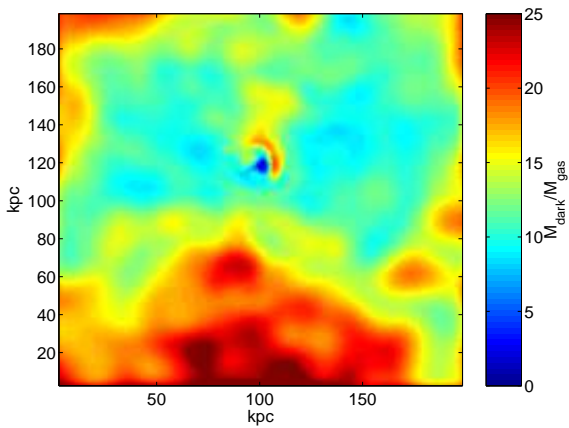}
\caption{
Mass ratio map in Simulation A. This frame corresponds to frame 7 in \fig{fig.Oscillations}.
}
\label{fig.MassRatioMap}
\end{figure}

The relative position of the gas and the dark matter haloes clarifies if the gas
is simply being dragged along with the dark matter or is also being displaced. 
\Fig{fig.MassRatioMap} maps the ratio of the dark matter mass to gas mass at a
time 200 Myr after closest approach in Simulation A (frame 7 of \fig{fig.Oscillations}).  The enhancement of the dark:gas ratio is evidence that it is the  dark matter distribution that is oscillating, bringing the gas with it.
However, there is a lag between the displacement of the  dark matter and the displacement of the gas.  The compression of the gas on the forward side leads to a gas pressure force which displaces the gas from the center of the potential well.  The displacement leads to adiabatic cooling of the gas on the leeward side and subsequent as the gas is pushed out of the potential well.

The sharp discontinuity of x-ray emission observed in cold-front clusters is an important feature of cold fronts that any theory must predict and explain.  The discontinuity is entirely attributable to a change in the gas density across the front, with the higher-density gas on the side of the front with the higher emission (contrary to the situation in which the emission excess is due to the presence of a shock).

\begin{figure}
\epsscale{\figscaleone}
\plotone{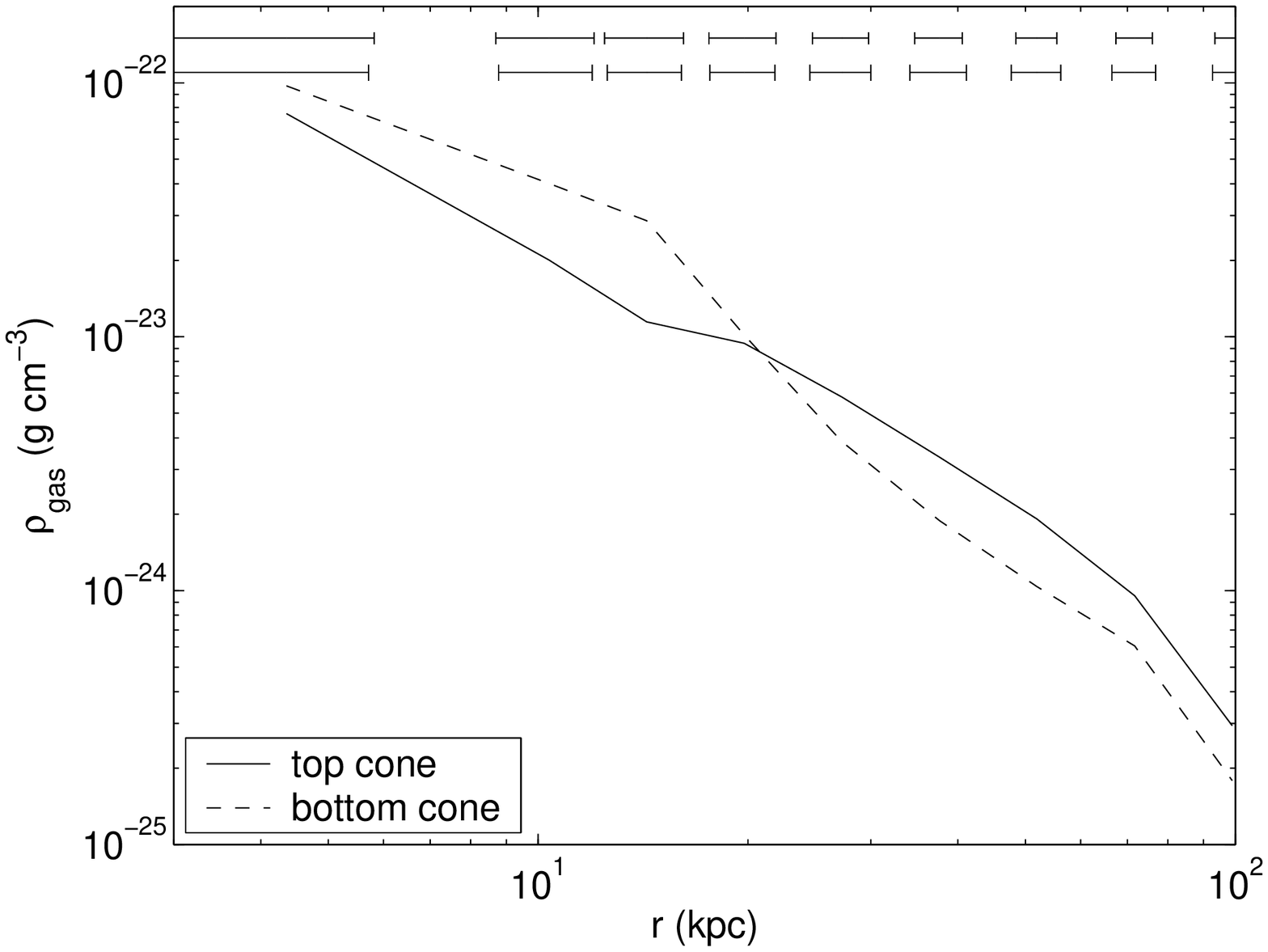}
\caption{
Density profiles for gas on opposite sides of the cluster center.  See the text for descriptions of the cones.  The mean gas smoothing radii are denoted by the horizontal lines.  The first row is for the top cone.  The second row for the bottom cone.  This frame corresponds to frame 6 in \fig{fig.Oscillations}. The density profiles do not have the sharp discontinuities the observations suggest, but the profiles are consistent with abrupt discontinuities softened by the SPH smoothing length.
}
\label{fig.InnerDensityProfiles}
\end{figure}

The simulations have yet to recreate the sharpness observed.
\Fig{fig.InnerDensityProfiles} illustrates the density profiles across the compression regions in a simulated cold front.  Two profiles are plotted, derived from the gas particles lying in conical regions extending outward and opposite from the center of the cluster.  The cluster is as in frame 6 in \fig{fig.Oscillations} with the cones having opening widths of $20\degr$.  The two cones extend above and below the cluster to contrast the density profiles in these regions.

At the resolution of the simulations, the simulated fronts are consistent with the sharpness of the discontinuities observed.
The simulation spatial resolution is provided by the SPH method.  The SPH smoothing length is set by the local number density of particles. The mean SPH smoothing radii for the particles as a function of radius is illustrated by the lines at the top of \fig{fig.InnerDensityProfiles}.  The mean SPH smoothing lengths in these regions are comparable to the width of the compression regions in the simulations.

It is as yet unknown whether increased spatial resolution provided by simulations with larger number of particles will lead to sharper density discontinuities.  According to the model presented in this paper, higher resolution simulations should produce sharper fronts.  The absence of sharper fronts in higher-resolution simulations will falsify the model.

\section{Comparison with observations}
\label{sec.Comparison}
The observational consequences of the model we have proposed permits morphological-based classification of cold-front clusters into oscillation and non-oscillation classes.  More than morphological data can be used in such a classification.  The velocity of the cold front is, on its own, a better discriminant.  However, determining the speed of a cold front from the x-ray data is not a trivial exercise.  It requires high quality data with spectroscopic information and assumptions about the relation between the temperature-density state of the gas and the velocity of the front.  The simple morphological rules described in this paper may be applied directly to the image data.

\begin{figure}
\epsscale{\figscaleone}
\plotone{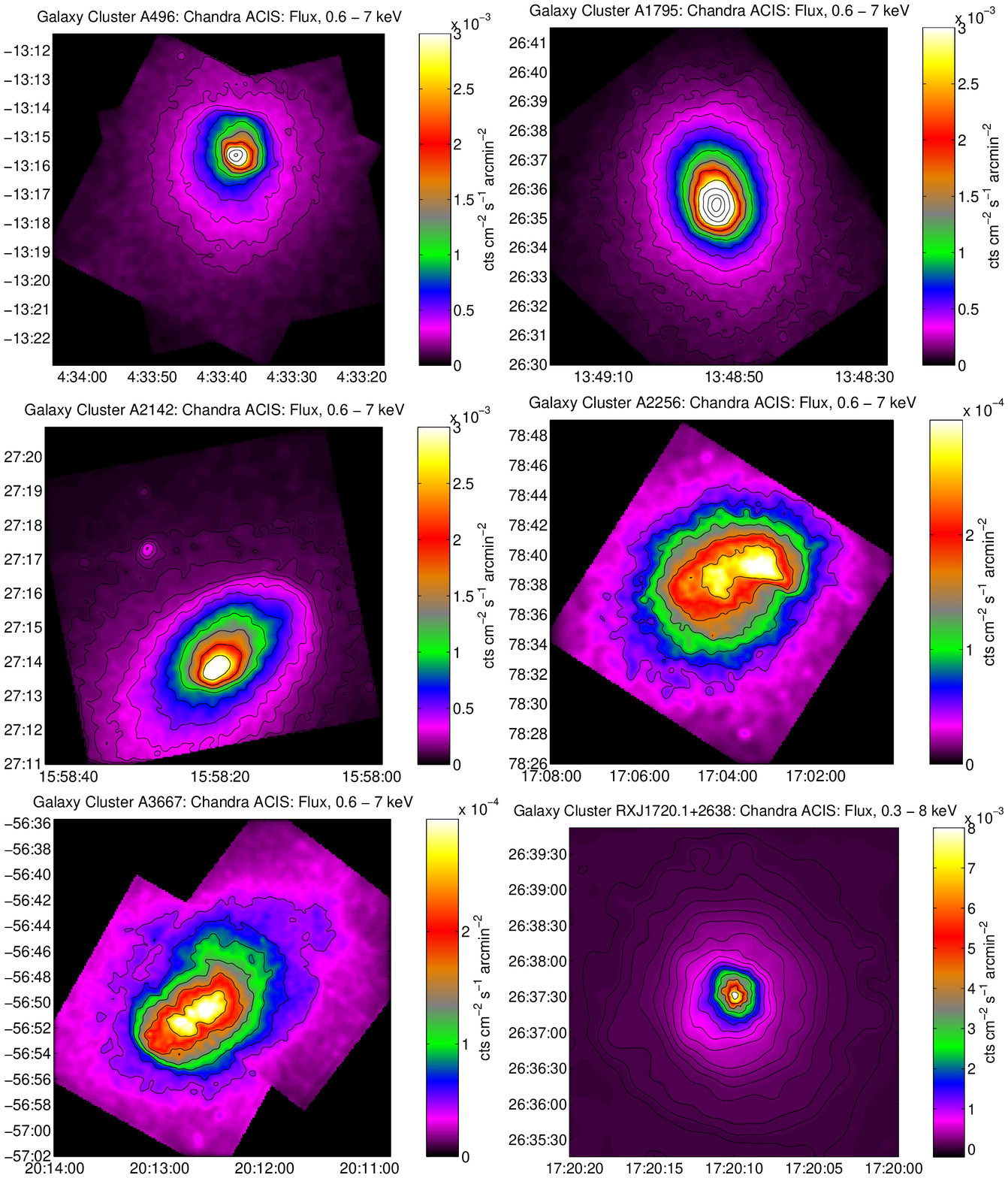}
\caption{
Galaxy clusters A496, A1795, A2142, A2256, A3667, and RXJ1720.1+2638.
The scales are, respectively,
$1\arcmin =  44 h_{65}^{-1} \kpc$,
$1\arcmin =  84 h_{65}^{-1} \kpc$,
$1\arcmin = 122 h_{65}^{-1} \kpc$,
$1\arcmin =  78 h_{65}^{-1} \kpc$,
$1\arcmin =  75 h_{65}^{-1} \kpc$, and
$1\arcmin = 220 h_{65}^{-1} \kpc$.
}
\label{fig.Clusters}
\end{figure}

In the literature, six clusters are described in sufficient detail to permit comparison with the model proposed here: A496, A1795, A2142, A2256, A3667, and RX J1720.1+2638.  We examine each of them here. \fig{fig.Clusters} shows the x-ray flux for the six clusters.  The data are from, respectively, ACIS-S3 chip over two observations, ACIS-S3 chip over three observations, ACIS-S3 chip over two observations, ACIS-I array (1 observation) and ACIS-S3 chip (3 observations), ACIS-I array over two observations, and ACIS-I array over one observation.  The images were created with an adaptive smoothing kernel. For 5 of the clusters, \citet{RB02} have published masses: A496 ($5.1\expd{14}h_{65}^{-1}\Msun$), A1795 ($11.9\expd{14}h_{65}^{-1}\Msun$), A2142 ($14.7\expd{14}h_{65}^{-1}\Msun$), A2256 $14.9\expd{14}h_{65}^{-1}\Msun$), and A3667 ($9.6\expd{14}h_{65}^{-1}\Msun$).  The range in cluster masses is similar to the range in the simulations.

A496 \citep{DW03} (\fig{fig.Clusters}) has a pair of non-axisymmetric isophotes discontinuities which are alternate and lie along an axis through the center of the cluster.  The isophotes are compressed $44 h_{65}^{-1} \kpc$ ($1\arcsec$)
SSE and $66 h_{65}^{-1} \kpc$ ($1.5\arcsec$) NNW of the cluster center. All three observational consequences of the oscillation model are satisfied.  
The model for the remnant class of objects, in which the group remnant is passing through the primary clusters halo, predicts a metallicity discontinuity at the front.  But the model for the oscillation class is inconsistent with a metallicity discontinuity since, unless there already exists an axisymmetric gradient, it is the in situ gas that is being compressed.  No metallicity discontinuity was found across the cold front in A496 by \citet{DW03}.  We classify A496 as an oscillation-type cold front.

\begin{figure}
\epsscale{\figscaleone}
\plotone{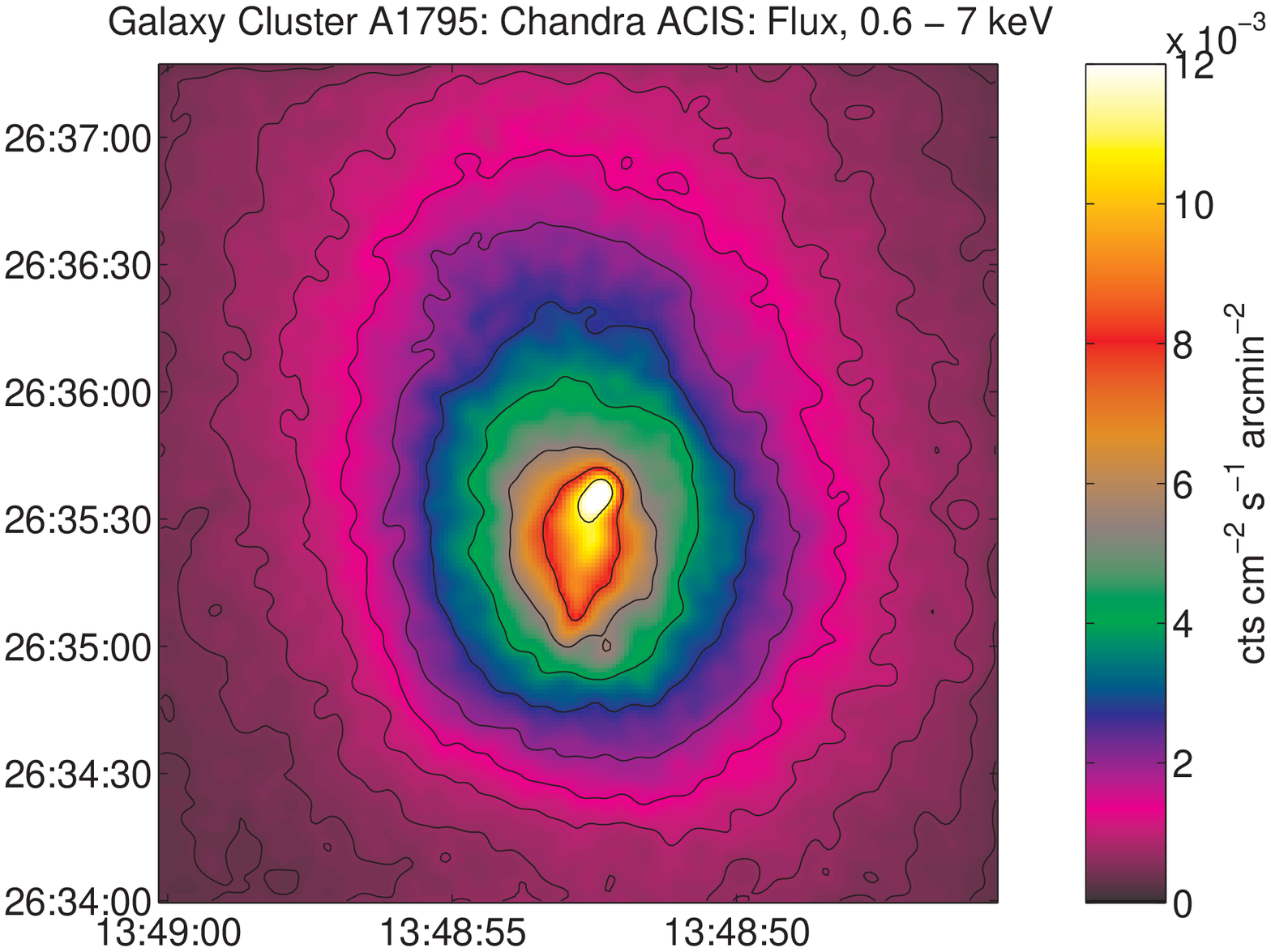}
\caption{
A detail of galaxy cluster A1795.
}
\label{fig.A1795_detail}
\end{figure}

A1795 \citep{MVM01} (\fig{fig.Clusters}) has a broad front at its southern edge
and a tighter front (\fig{fig.A1795_detail}) at the northern side of the core in what is described as a filament. The isophotes are compressed $11 h_{65}^{-1} \kpc$ ($0.13\arcsec$) N and $84 h_{65}^{-1} \kpc$ ($1\arcsec$) S of the cluster center. The fronts lie along the north-south axis.  Hence, the three observational requirements for an oscillation class cold front are satisfied and we classify A1795 as such.

A2142 \citep{Markevitch00}, x-ray flux map shown in \fig{fig.Clusters}, exhibits a clear compression of its isophotes $87 h_{65}^{-1} \kpc$ ($0.7\arcsec$) SE of the cluster center and at larger radii $300 h_{65}^{-1} \kpc$ ($2.5\arcsec$) NW of the center. The compression regions are alternate and lie along an axis passing through the cluster center. We classify A2142 as containing an oscillation-class cold front.

A2256 \citep{SMMV02} (\fig{fig.Clusters}) is revealed in the x-ray band to contain an irregularly shaped knot of cool dense gas located to the west of the cluster center.  The sharpest edge of the cool gas faces SSE.  There is only one edge.  The gradient of the edge is not along an axis through the cluster center.  A2256 does not contain an oscillation-type cold front.  We classify A2256 as a merger-class front.

A3667 \citep{MSV99, VMM01b, MFV02}, shown in \fig{fig.Clusters}, has a single cold front $410 h_{65}^{-1} \kpc$ ($5.5\arcsec$) SE of the cluster center.  It also contains a filament of gas to the NE of the cluster center.  Though the gradient in the region of the front does point through the cluster center, the single front and the filament are more consistent with an ongoing major merger.  Simulations reproduce the morphology with a massive head-on merger \citep{RBS99}.  We classify A3667 as a merger-class front.

RX J1720.1+2638 \citep{Mazzotta01} is an almost relaxed cluster (see \fig{fig.Clusters}).  There is a clear compression of the isophotes at
three locations: two to the SE at $44 h_{65}^{-1} \kpc$ ($0.2\arcmin$) and $240 h_{65}^{-1} \kpc$ ($1.1\arcmin$) from the center and one $88 h_{65}^{-1} \kpc$ ($0.4\arcmin$) to the NW.  All are along a position angle of $140\degr$, compressed perpendicular to the position angle.  The position and alignment of the compression regions is consistent with oscillation-class fronts.

Of the six clusters reported in the literature to have cold fronts, four (A496, RX J1720.1+2638, A2142, and A1795) are consistent in morphology with oscillation-class cold fronts.  Two are inconsistent (A2256 and A3667) and we
classify without further examination as having remnant-class cold fronts.  Six clusters are too few to permit a satisfying statistical sample.  However, if $\twothirds$ of clusters have cold fronts \citep{MVF02},  while less than $\onethird$ are expected to have merger remnants within their haloes \citep{FSNY02}, then more than half the cold front clusters must originate from some process other than remnant group haloes.

\begin{figure}
\epsscale{\figscaleone}
\plotone{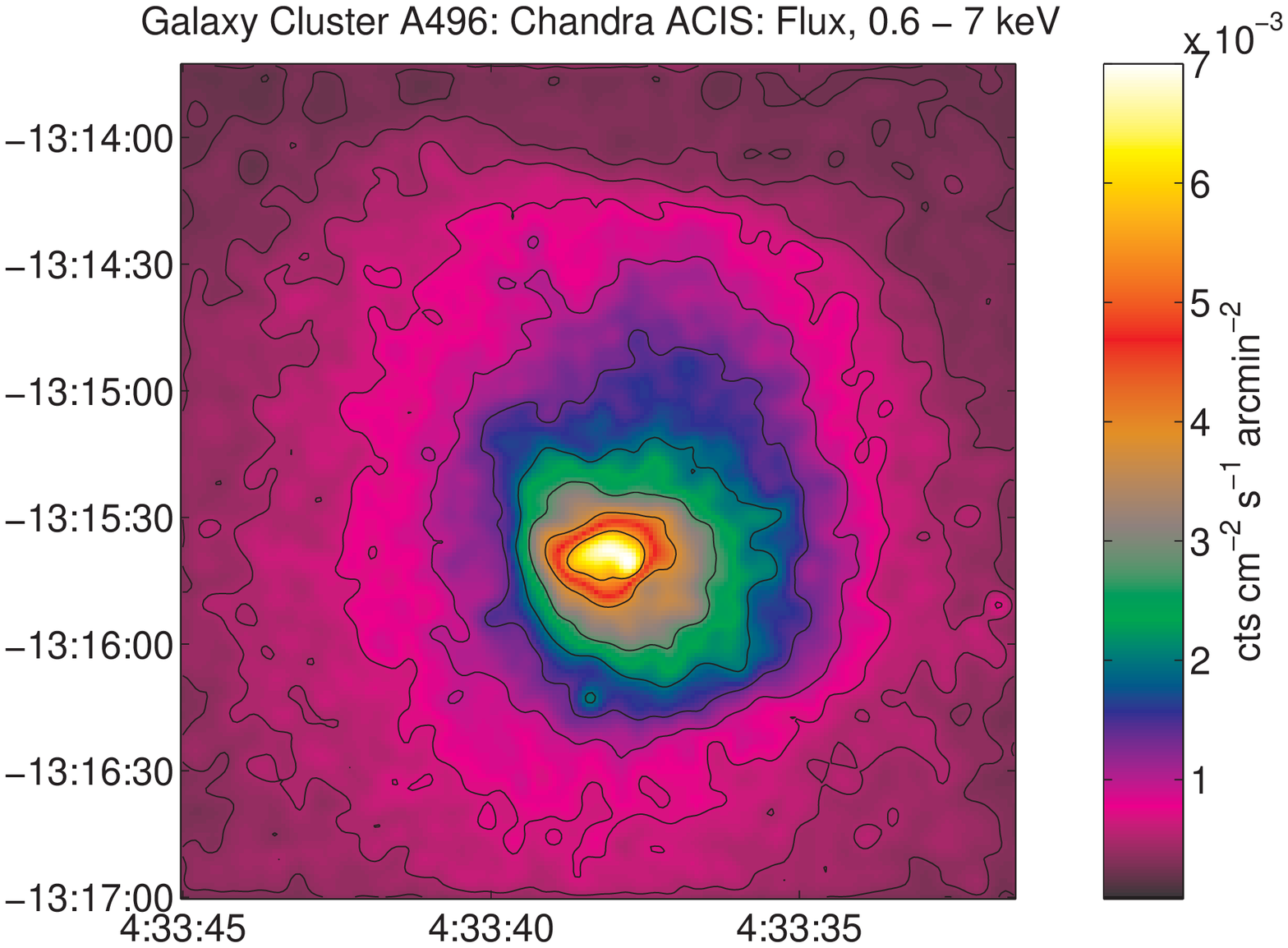}
\caption{Detail of galaxy cluster A496.}
\label{fig.A496_detail}
\end{figure}

In the simulations, the compression regions were found over a range of scales from 8 to 80 kpc.  The observed compression regions in the clusters classified as oscillation type were over a similar range in magnitude but overall larger scale: 11 to $300 h_{65}^{-1} \kpc$.  The smallest oscillation feature in the simulations (8 kpc) would correspond to only $10\arcsec$ in the \chandra\ observation of the nearest cluster (A496).  \chandra\ is not limited by resolution in resolving such small features, but it is limited by counts.  \fig{fig.A496_detail} reveals the inner $3\arcmin$ region of A496.  There is a region of compressed isophotes $15 h_{65}^{-1} \kpc$ ($21\arcsec$) E of the x-ray peak.  The region is not collinear with the other two regions of compression to the north and south of the cluster center, but the gradient vector is parallel with a line passing through the x-ray peak, raising the possibility that A496 is undergoing two concurrent oscillations.

\section{Energy in the oscillations and cooling flows}
\label{sec.Energy}
The temperature of the gas in the centers of cooling flow clusters are consistently measured to not fall below $\onethird T_{virial}$ \citep{Fabian02}. This runs contrary to expectation; the radiative efficiency of cool dense gas does not fall off until below $\sim 10^4 \K$.
To prevent gas from cooling catastrophically requires an energy injection method that is both plentiful and not instantaneous.  \citet{Fabian02}
notes a required rate of energy injection of up to $10^{45}\erg \sec^{-1}$.  The kinetic energy in the oscillations can both provide a source of energy and spread this energy injection over a period of time.  But is it enough?

\begin{figure}
\epsscale{\figscaleone}
\plotone{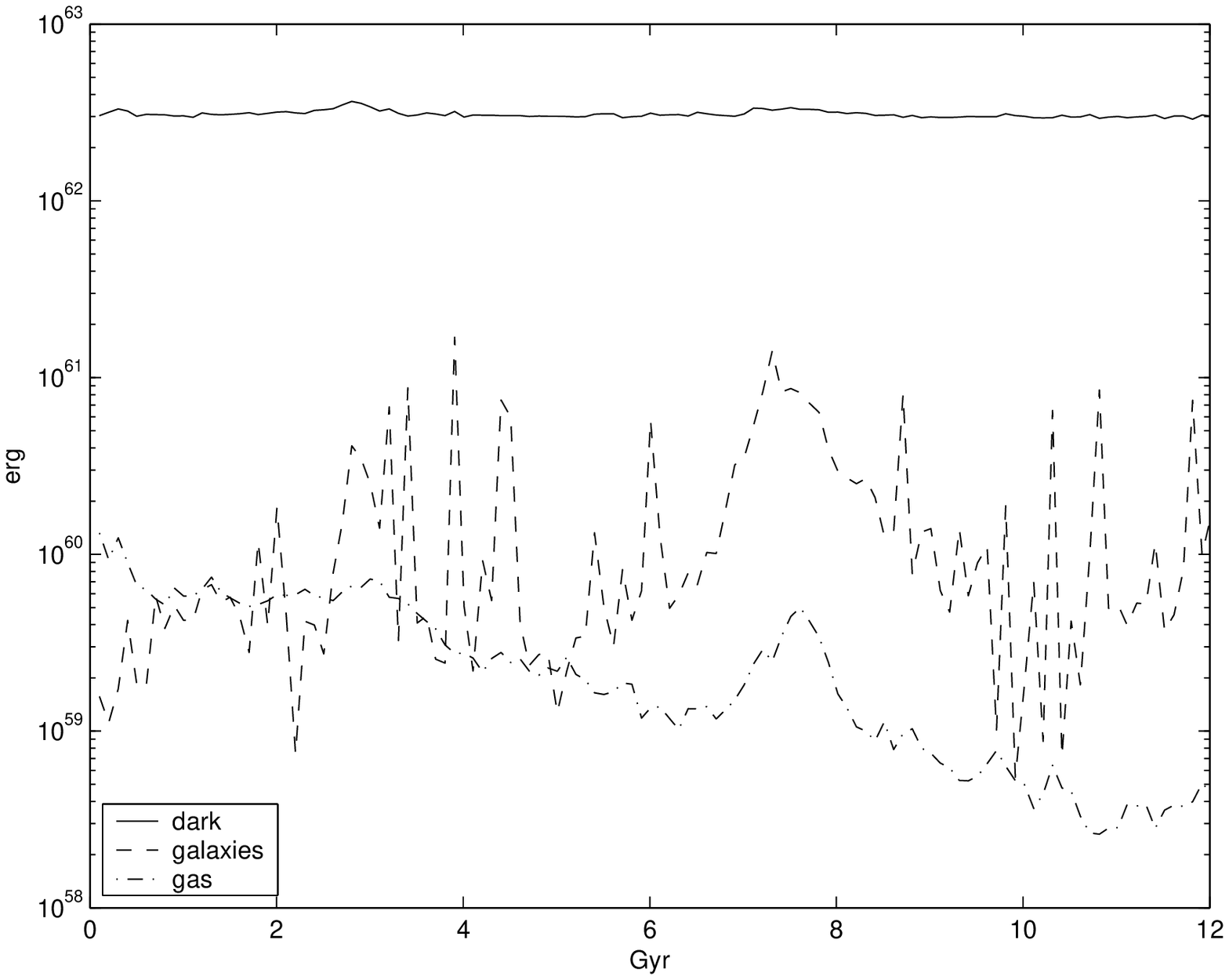}
\caption{Kinetic energy in the dark matter, galaxies, and gas for Simulation B.}
\label{fig.Energy}
\end{figure}

An examination of the energy stored as kinetic energy in the three phases demonstrates that oscillations could be the unknown source of power that regulates the central temperature in cooling flow clusters provided there exists a mechanism to transfer kinetic energy from the galaxies.  The cluster in Simulation B has a mass of $4.2\expd{14}\Msun$ which is not particularly massive. Clusters with similar masses are observed to have luminosities of about $10^{44}\erg \sec^{-1}$.  
\fig{fig.Energy} graphs the kinetic energy stored in the dark matter, galaxies, and gas in Simulation B.  Closest approach occurs at $7.2 \Ga$.  For the gas, a minimum of $4\expd{59} \erg$ is stored as kinetic energy in the gas phase.  The gas disperses the energy in 700 Myr for a mean heating rate of $1.8\expd{43} \erg \sec^{-1}$.  This is one fifth of the required injection rate.  However, the other two phases: galaxies and dark matter, store substantially more energy. The relative motions of galaxies store a minimum kinetic energy of $1\expd{61} \erg$ and disperse this energy over a longer time-scale, $\sim 2.4 \Ga$, for a mean injection rate of $1.3\expd{44} \erg \sec^{-1}$ which is of the same order of magnitude as the luminosity.  The largest reservoir of kinetic energy is the dark matter.  The dark matter receives at least $3.6\expd{61} \erg$ from the interaction which it dissipates in $1.2 \Ga$ for a mean dissipation rate of $1.0\expd{45} \erg \sec^{-1}$.  Though this is the well above the amount of power required, there are no efficient means for transferring energy between the two phases.  If a mechanism can be found to efficiently transfer galactic
kinetic energy into gas thermal energy, the energetics of the system are sufficient to maintain the minimum temperature in the centers of cooling flow clusters.

\section{Summary}
\label{sec.Summary}
The cluster morphology observed in some clusters with cold fronts has been shown to arise from oscillations of the cluster core induced by gravitational interactions with subclusters.  Numerical simulations with hydrodynamics give us a list of morphological signatures of these oscillations:
\begin{enumerate}
 \item There can be one, two, or three regions of compressed isodensity contours.
 \item The regions of compression must be alternate of each other and lie on an axis which passes through the center of the cluster.
 \item The gradients of compressed isodensity must point through the center of the cluster.
\end{enumerate}
We associate the compressed isodensities with cold fronts and show that 4 of 6 clusters with cold fronts examined exhibit these signatures.  The other two clusters which do not exhibit these features are likely undergoing a merger of a subcluster with the cold gas coming from the remnant of the subcluster.

That the oscillations create cold-front morphologies solves the discrepancy between the number of clusters with cold-fronts observed and the number expected if cold-fronts were due only to the remnants of merged subclusters.  The oscillation class has two major advantages to explaining the increased frequency:
\begin{enumerate}
\item The oscillations are able to persist long after the passage of the
	merging structure.
\item Major mergers are not required to instigate oscillations.  More
	common high-mass-ratio interactions are sufficient.
\end{enumerate}

The energy contained in the kinetic energy of the oscillating intracluster medium is insufficient to offset that lost by radiative cooling and hence does not solve the cooling flow problem.  However, the kinetic energy carried by the galaxies in the oscillation is of the correct magnitude.  In either case, the duration over which the energy is released to the cluster intracluster medium is $\sim 100$ Myr to $\sim 1000$ Myr, providing a steady source of energy.

\bibliographystyle{apj}
\bibliography{apj-jour,biblio}

\begin{thebibliography}{20}
\expandafter\ifx\csname natexlab\endcsname\relax\def\natexlab#1{#1}\fi

\bibitem[{{Bialek} {et~al.}(2002){Bialek}, {Evrard}, \& {Mohr}}]{BEM02}
{Bialek}, J.~J., {Evrard}, A.~E., \& {Mohr}, J.~J. 2002, \apjl, 578, L9

\bibitem[{{Buote} \& {Tsai}(1996)}]{BT96}
{Buote}, D.~A. \& {Tsai}, J.~C. 1996, \apj, 458, 27

\bibitem[{Couchman {et~al.}(1995)Couchman, Thomas, \& Pearce}]{CTP95}
Couchman, H. M.~P., Thomas, P.~A., \& Pearce, F.~R. 1995, \apj, 452, 797

\bibitem[{{De Grandi} \& {Molendi}(2001)}]{DM01}
{De Grandi}, S. \& {Molendi}, S. 2001, \apj, 551, 153

\bibitem[{{Dupke} \& {White}(2003)}]{DW03}
{Dupke}, R. \& {White}, R.~E. 2003, \apjl, 583, L13

\bibitem[{{Fabian} {et~al.}(2002){Fabian}, {Allen}, {Crawford}, {Johnstone},
  {Morris}, {Sanders}, \& {Schmidt}}]{Fabian02}
{Fabian}, A.~C., {Allen}, S.~W., {Crawford}, C.~S., {Johnstone}, R.~M.,
  {Morris}, R.~G., {Sanders}, J.~S., \& {Schmidt}, R.~W. 2002, \mnras, 332, L50

\bibitem[{{Fujita} {et~al.}(2002){Fujita}, {Sarazin}, {Nagashima}, \&
  {Yano}}]{FSNY02}
{Fujita}, Y., {Sarazin}, C.~L., {Nagashima}, M., \& {Yano}, T. 2002, \apj, 577,
  11

\bibitem[{{Markevitch} {et~al.}(2000){Markevitch}, {Ponman}, {Nulsen}, {Bautz},
  {Burke}, {David}, {Davis}, {Donnelly}, {Forman}, {Jones}, {Kaastra},
  {Kellogg}, {Kim}, {Kolodziejczak}, {Mazzotta}, {Pagliaro}, {Patel}, {Van
  Speybroeck}, {Vikhlinin}, {Vrtilek}, {Wise}, \& {Zhao}}]{Markevitch00}
{Markevitch}, M., {Ponman}, T.~J., {Nulsen}, P.~E.~J., {Bautz}, M.~W., {Burke},
  D.~J., {David}, L.~P., {Davis}, D., {Donnelly}, R.~H., {Forman}, W.~R.,
  {Jones}, C., {Kaastra}, J., {Kellogg}, E., {Kim}, D.-W., {Kolodziejczak}, J.,
  {Mazzotta}, P., {Pagliaro}, A., {Patel}, S., {Van Speybroeck}, L.,
  {Vikhlinin}, A., {Vrtilek}, J., {Wise}, M., \& {Zhao}, P. 2000, \apj, 541,
  542

\bibitem[{{Markevitch} {et~al.}(1999){Markevitch}, {Sarazin}, \&
  {Vikhlinin}}]{MSV99}
{Markevitch}, M., {Sarazin}, C.~L., \& {Vikhlinin}, A. 1999, \apj, 521, 526

\bibitem[{Markevitch {et~al.}(2002)Markevitch, Vikhlinin, \& Forman}]{MVF02}
Markevitch, M., Vikhlinin, A., \& Forman, W.~R. 2002, astro-ph/0208208v1

\bibitem[{{Markevitch} {et~al.}(2001){Markevitch}, {Vikhlinin}, \&
  {Mazzotta}}]{MVM01}
{Markevitch}, M., {Vikhlinin}, A., \& {Mazzotta}, P. 2001, \apjl, 562, L153

\bibitem[{{Mazzotta} {et~al.}(2002){Mazzotta}, {Fusco-Femiano}, \&
  {Vikhlinin}}]{MFV02}
{Mazzotta}, P., {Fusco-Femiano}, R., \& {Vikhlinin}, A. 2002, \apjl, 569, L31

\bibitem[{{Mazzotta} {et~al.}(2001){Mazzotta}, {Markevitch}, {Vikhlinin},
  {Forman}, {David}, \& {VanSpeybroeck}}]{Mazzotta01}
{Mazzotta}, P., {Markevitch}, M., {Vikhlinin}, A., {Forman}, W.~R., {David},
  L.~P., \& {VanSpeybroeck}, L. 2001, \apj, 555, 205

\bibitem[{{Mohr} {et~al.}(1995){Mohr}, {Evrard}, {Fabricant}, \&
  {Geller}}]{MEFG95}
{Mohr}, J.~J., {Evrard}, A.~E., {Fabricant}, D.~G., \& {Geller}, M.~J. 1995,
  \apj, 447, 8

\bibitem[{{Mohr} {et~al.}(1993){Mohr}, {Fabricant}, \& {Geller}}]{MFG93}
{Mohr}, J.~J., {Fabricant}, D.~G., \& {Geller}, M.~J. 1993, \apj, 413, 492

\bibitem[{{Nagai} \& {Kravtsov}(2003)}]{NK03}
{Nagai}, D. \& {Kravtsov}, A.~V. 2003, \apj, 587, 514

\bibitem[{{Reiprich} \& {B{\" o}hringer}(2002)}]{RB02}
{Reiprich}, T.~H. \& {B{\" o}hringer}, H. 2002, \apj, 567, 716

\bibitem[{{Roettiger} {et~al.}(1999){Roettiger}, {Burns}, \& {Stone}}]{RBS99}
{Roettiger}, K., {Burns}, J.~O., \& {Stone}, J.~M. 1999, \apj, 518, 603

\bibitem[{{Sun} {et~al.}(2002){Sun}, {Murray}, {Markevitch}, \&
  {Vikhlinin}}]{SMMV02}
{Sun}, M., {Murray}, S.~S., {Markevitch}, M., \& {Vikhlinin}, A. 2002, \apj,
  565, 867

\bibitem[{{Vikhlinin} {et~al.}(2001){Vikhlinin}, {Markevitch}, \&
  {Murray}}]{VMM01b}
{Vikhlinin}, A., {Markevitch}, M., \& {Murray}, S.~S. 2001, \apj, 551, 160

\end{thebibliography}

\end{document}